\documentclass[pra,twocolumn,showpacs]{revtex4-1}
\usepackage{graphicx,epstopdf}
\usepackage{amsmath}
\usepackage{amssymb}
\usepackage{epsfig}
\usepackage{epstopdf}
\usepackage{subfigure}
\usepackage{enumerate}

\usepackage[colorlinks=true, citecolor=blue, urlcolor=blue ]{hyperref}

\begin{document}
\title{Growth of genuine multipartite entanglement  in random unitary circuits
}
\author{Anindita Bera\(^{1}\) and Sudipto Singha Roy\(^{2}\)}
\affiliation{\(^1\)Racah Institute of Physics, The Hebrew University of Jerusalem, Jerusalem 91 90401, Givat Ram, Israel\\
\(^2\)\emph{Instituto de F{\'i}sica T{\'e}orica UAM/CSIC,  C/ Nicol{\'a}s Cabrera 13-15, Cantoblanco, 28049 Madrid, Spain}}
\date{\today}
\begin{abstract}
We study the growth of genuine multipartite entanglement in random quantum circuit models,  which  include random unitary circuit models  and the random Clifford circuit.  We find that for the random Clifford circuit, the growth of multipartite entanglement remains slower in comparison to the random unitary case. However, the final saturation value of multipartite entanglement is almost the same in both cases.    The behavior is then compared to the genuine multipartite entanglement obtained in random matrix product states with a moderately high bond dimension. We then relate the behavior of multipartite entanglement to other global properties of the system, viz. the delocalization of the many-body wavefunctions in Hilbert space. Along with this, we analyze the robustness of such  highly entangled quantum states obtained through random unitary dynamics under weak measurements.
\end{abstract}
\maketitle

\section{Introduction}
 Over the years,  alongside quantum Hamiltonian systems,  an important area that has gained  much attention   is the study of quantum properties related to the quantum  random unitary circuit models~\cite{random_unitary0,random_unitary1,random_unitary2,random_unitary3,random_unitary4,random_unitary5,random_unitary6,random_unitary7,random_unitary8,random_unitary9,random_unitary10, cross_entropy_Supremacy1,cross_entropy_Supremacy2,cross_entropy_Supremacy3,new_reference1}. Even though such a   model is relatively less-structured than any generic Hamiltonian system, as it  retains only  two fundamental features of any realistic physical system, namely, unitarity and spatial locality, several studies reveal that it is comprised of many rich quantum properties.   It has been reported that quantum entanglement growth in these systems exhibits a certain universal structure~\cite{random_unitary1}.  In particular, the critical exponents for the entanglement growth are similar to those of the Kardar-Parisi-Zhang (KPZ) equation,  which has a wide range of applicability in non-equilibrium statistical mechanics \cite{surface_growth}. Moreover, the exact expression of the rate of entanglement growth commonly known as entanglement speed is obtained for these models by  rewriting the dynamics of purity  as a classical Markov process and mapping it to solvable spin models \cite{new_reference1}.
In addition to this, both exact results and coarse-grained descriptions have been provided for the spreading of quantum operators under random quantum circuit dynamics \cite{random_unitary2,random_unitary4}. Along with this,  a new class of dynamical behavior has been explored, when the circuit is constantly monitored through quantum measurements~\cite{random_unitary6,random_unitary7,
random_unitary8,random_unitary9,random_unitary10}. 
 Interestingly, it has been reported that the entanglement growth of an initial product state under such random unitary dynamics, undergoes a continuous transition from the volume-law to the area-law, when it is monitored with a  particular strength of the measurement~\cite{random_unitary6,random_unitary7,random_unitary8,random_unitary9,random_unitary10}.

Apart from entanglement studies,  random quantum circuits have also been used to demonstrate quantum advantage through the task of sampling from the output distributions of the models~\cite{cross_entropy_Supremacy1,cross_entropy_Supremacy2,cross_entropy_Supremacy3}. In general, the accomplishment of such a task classically requires a direct numerical simulation of the circuit, with computational cost exponential in the number of qubits~\cite{cross_entropy_Supremacy1,cross_entropy_Supremacy2}.  Very recently, experimental validation of the same has also been reported~\cite{cross_entropy_Supremacy3}.   Along with this, random quantum circuits composed of nearest neighbor two-qubit gates have been proven to form an approximate unitary $t$-design \cite{new_reference2,new_reference3,new_reference4}. Pseudorandomness in the form of unitary $t$-design has emerged as a promising approach for experimental realization of random unitary circuits. Despite its mathematical simplicity, an exact experimental realization of random quantum circuits demands an extremely long time and is often unfeasible in many-body systems. In this regard, the scheme of unitary $t$-design  provides finite-degree approximation of Haar random unitaries and has already been implemented experimentally in small systems \cite{unitary_t_design_exp1,unitary_t_design_exp2,unitary_t_design_exp3,unitary_t_design_exp4,unitary_t_design_exp5}. All this progress has significantly increased the interest of the community to explore several other quantum properties related to random circuit models.

To date, among the works related to random unitary circuit models, the study of local or bipartite quantum properties has received most of the attention. However, along with those properties, an important case to explore is the global quantum properties of these models, which are complex, although fundamentally interesting. In particular, multipartite entanglement is one such important property which is also considered to be a  potential resource in many quantum information and computation protocols~\cite{multiparty_QIP1,multiparty_QIP2,multiparty_QIP3,
multiparty_QIP4,multiparty_QIP5,multiparty_QIP6,error_corr0,error_corr1}. In recent years, several studies endorse the fact that along with bipartite entanglement, multipartite entanglement can also faithfully detect quantum phase transitions in several quantum many-body systems~\cite{ME_QTP1,ME_QTP2,ME_QTP3,ME_QTP4,ME_QTP5,ME_QTP6,ME_QTP7,multiparty-more1}. Moreover, using recent technologies, experimental realizations and manipulation of multipartite entangled states have also been reported in atomic, ion-trap and optical settings~\cite{multiparty_ent1, multiparty_ent2,multiparty_ent4,multiparty_ent6,multiparty_ent7,multiparty_ent8}.

In this article, we address this void and aim at studying the global quantum properties of random unitary circuits. In particular, we look at the multipartite entanglement properties of the quantum state obtained at each iteration of a random unitary circuit and relate it to other physical properties of the system.  We observe that an initial product state when subjected to a random unitary circuit, acquires a substantial amount of genuine multipartite entanglement even for a few iterations of the circuit, and at large iteration time, eventually saturates to a very high value, close to the maximum possible value in qubit systems. Additionally, we find that the growth rate of genuine multipartite entanglement has a dependence on the range of interactions. For instance,  we observe that though the saturation value of multipartite entanglement in random unitary circuit models comprised of quasi long-range and long-range unitaries remain same as that obtained for short-range circuits, as all of them are ergodic and converge to the same Haar measure,  saturation in these cases is attained with a   faster rate than the short-range one. Along with this,  we also present the results obtained from the analysis of multipartite entanglement properties of another form of a random quantum circuit, which is composed of structurally different elementary gates, namely, the random Clifford circuit \cite{Clifford1,Clifford2}. We note that in this case, the rate of growth of multipartite entanglement remains slow in comparison to random unitary circuits and saturation occurs at a much higher value of the iteration step of the circuit. However, the final saturation value of multipartite entanglement is found to be almost the same as that obtained for the random unitary case. We then compare the behavior of multipartite entanglement obtained for the random unitary circuit to that obtained for random matrix product states (RMPS) \cite{Random_MPS}.  This provides us a framework to characterize the complexity of the random state generated at each iteration of the circuit in terms of the bond dimension of the random matrix product states.

Once the multipartite entanglement properties of the circuits are fully characterized, we next relate that to other global quantum properties of these models. In particular, we study the delocalization of the initial wavefunction when it is subjected to the quantum dynamics under the random quantum circuits.     In general, for any generic quantum many-body wavefunction, the relation between its spread in the Hilbert space or delocalization and its global entanglement content is not obvious. In this respect,   for some specific quantum-many body states, close resemblance of the behavior of bipartite entanglement and localization have been reported in many earlier works~\cite{IPR_EE001,IPR_EE002,IPR_EE1, IPR_EE2, IPR_EE3}.
To find whether any such relationship exists in case of random dynamics we consider in this work, we compute the inverse-participation-ratio (IPR) \cite{IPR_new,IPR_EE0} in local basis,  for both random unitary circuits and random Clifford circuit. We note that the behavior of  IPR in both the circuits remain very much akin to their global entanglement properties.  Therefore, we  argue that in the random quantum  circuits we have considered in our work, the spreading of the quantum many-body wavefunctions in Hilbert space and the growth of multipartite entanglement have close correspondence.

Finally, we analyze the robustness of multipartite entanglement of the random quantum state generated for a large number of circuit iterations, when it is monitored through non-projective or unsharp or weak measurements \cite{weak_measure1,weak_measure2, weak_measure3}, which are a special subset of positive-operator-valued-measurements (POVMs).
We report that the random quantum state sustains a non-zero amount of global entanglement, even for high values of the measurement strength. Interestingly, the decay pattern of global entanglement with the measurement strength becomes almost similar to that obtained for an $N$-qubit GHZ state.

Therefore,  our work sheds light on fundamental as well as application-based aspects of global quantum properties of the random unitary circuit models that have not been addressed in previous works. We argue that random quantum circuits,  even though result in less-structured quantum dynamics than a generic Hamiltonian system, generate a high amount of global entanglement, which can be a promising scheme for efficient generation and control of the multipartite entangled state. Along with this, the robustness property of the multipartite entanglement under weak measurements opens up the possibility to use it as a potential resource in quantum information and computation tasks which are accomplished exploiting multipartite entanglement~\cite{multiparty_QIP1,multiparty_QIP2,multiparty_QIP3,
multiparty_QIP4,multiparty_QIP5,multiparty_QIP6,error_corr0,error_corr1}.

We arrange the paper in the following way. In Sec.~\ref{model}, we describe the random unitary circuit that we consider in our work. In Sec.~\ref{GGM_def}, we briefly discuss the measure of genuine multipartite entanglement. Next, in Sec.~\ref{ggm_growth}, we demonstrate the growth of genuine multipartite entanglement with each iteration step of the random unitary circuits and compare it with that obtained for random Clifford  circuit. In  Sec.~\ref{GGM_MPS}, we compare the behavior of multipartite entanglement obtained for the random state generated through  random unitary circuit to that obtained for a random matrix product state.  A comparison between two global properties of the circuits, namely, the spread of wavefunctions in Hilbert space and the multipartite entanglement are made in Sec.~\ref{GGM_IPR}. Next, in Sec.~\ref{Weak_measurement}, we discuss the robustness of the multipartite entanglement generated in the random unitary circuit under the effect of weak measurements. Finally, we conclude in Sec.~\ref{conclusion}.\\

\begin{figure}[h]
\includegraphics[width=8.2cm]{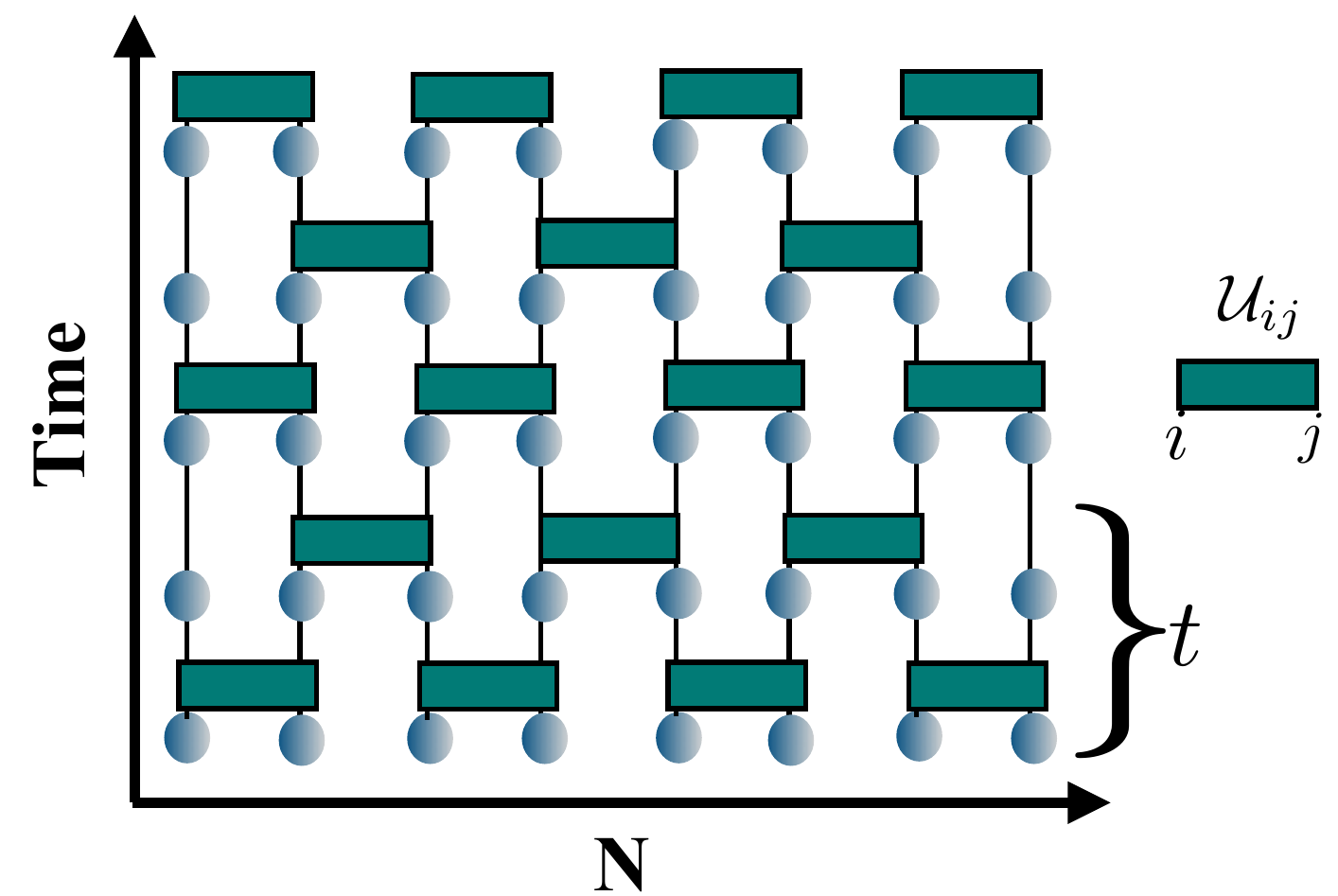}
\caption{Schematic representation of the random unitary circuit. The solid circles indicate the lattice sites and the rectangular boxes represent the random unitaries $\mathcal{U}_{ij}$ acting on sites  $i$ and $j$. The abscissa and ordinate correspond to the number of sites $N$ of the circuit and the evolution time respectively. A complete iteration is denoted by $t$.}
\label{schematic}
\end{figure} 

\section{The model}
\label{model}
Let us briefly discuss the random unitary circuit as depicted in  Fig.~\ref{schematic}. In this circuit, we first apply random unitaries  $\mathcal{U}_{ij}$, generated independently through Haar measure, on the nearest neighbor sites $(i,j)$, i.e., on the sites $(1,2), (3,4), (5,6), \dots,(N-1, N)$, where $N$ is the total number of sites. In the next step, we apply the unitaries on the remaining nearest-neighbor pairs of sites, i.e., $(2,3), (4,5), (6,7), \dots, (N-2, N-1)$. This completes a full iteration, denoted by $t$. The number of unitaries acting at each iteration is $N-1$.  As the interactions between spins or qubits are taken to be random in both space and time, it becomes a less-structured model than any generic Hamiltonian system. Note here that throughout the paper, we mainly consider the random states generated through the short-range random unitary circuits discussed above. However,  in our work, we also study some of the variants of random unitary circuits, which we discuss in detail  in Sec. \ref{ggm_growth}.

\section{Genuine multipartite entanglement and its measure}
\label{GGM_def}
In this section, we briefly introduce the measure of genuine multipartite entanglement that we consider in our work. An $N$-party pure quantum state $|\Psi\rangle_N$ in the tensor product Hilbert space $\mathcal{H}_1 \otimes \mathcal{H}_2 \otimes \dots \mathcal{H}_{N}$  \cite{zanardi}  is said to be genuinely multipartite entangled if it cannot be written as a product in any possible bipartitions of the state~\cite{ggm1,ggm2,ggm3,ggm4,ggm5}. An example of such quantum state is the $N$-party GHZ state, given by $|\psi\rangle_{\mbox{\small{GHZ}}}=\frac{1}{\sqrt{2}}(|0\rangle^{\otimes^N}+|1\rangle^{\otimes^N})$.  In order to quantify the genuine multipartite entanglement of any pure  quantum state $|\Psi\rangle_N$, we consider a computable measure known as the generalized geometric measure (GGM)~\cite{ggm3,ggm4,ggm5}. It is defined as an optimized  distance of the given quantum state, $|\Psi\rangle_N$, from the set of all states that are not genuinely multipartite entangled. This can be mathematically expressed as
\begin{equation}
\mathcal{G}(|\Psi\rangle_N)=1-\zeta_{\max}^2(|\Psi\rangle_N),
\end{equation}
where $\zeta_{\max} (|\Psi\rangle_N ) = \max | \langle \eta|\Psi\rangle_N |$, with the maximization being carried out over all  pure $N$-party quantum state $|\eta\rangle\in \mathcal{H}_1 \otimes \mathcal{H}_2 \otimes \dots \mathcal{H}_{N}$, which are  not genuinely multipartite entangled. Further simplification of the above equation leads to an equivalent expression, given  by
\begin{equation}\mathcal{G} (|\Psi \rangle_N ) =  1 - \max \{\lambda_{ A : B} | A \cup  B = \{1,2,\ldots,N\},~A \cap B = \emptyset\},
\label{GGM}
\end{equation}
where \(\lambda_{A:B}\)  is the largest eigenvalue of the reduced density matrix $\rho_A$ or $\rho_B$ of $|\Psi\rangle_N$. For the qubit system, the value of $\mathcal{G}$ lies within the range  $0\leq \mathcal{G} \leq 0.5$.

\begin{figure}[t]
\includegraphics[width=8.8cm]{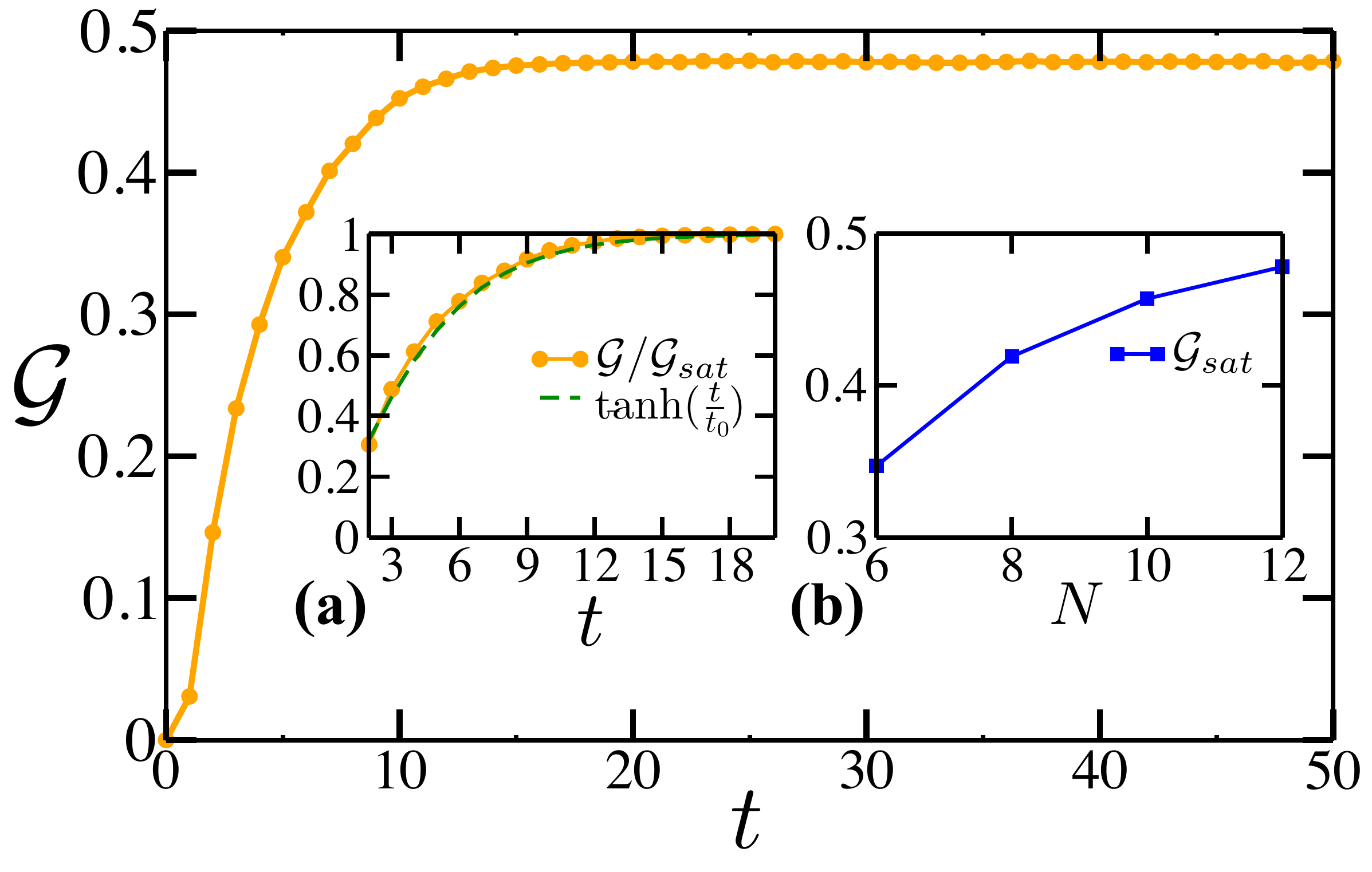}
\caption{The growth of $\mathcal{G}$ with iteration number $t$. At each iteration, the averaging of $\mathcal{G}$ is performed over $2\times 10^2$ random realizations of the unitaries. Here $N=12$. Inset (a) shows the fitting of  $\mathcal{G}/\mathcal{G}_{sat}$ vs. $t$  profile which is close to the function  $ \tanh(\frac{t}{t_0})$ (see Appendix \ref{appA}). Inset (b) exhibits the scaling of $\mathcal{G}_{sat}$ with $N$.}
\label{GGM_plot}
\end{figure}

\section{Growth of multipartite entanglement}
\label{ggm_growth}

We are now equipped with the necessary tools to study the global entanglement properties of an initial product state when it is iteratively subjected to the random quantum circuit described in Fig.~\ref{schematic}. Towards this aim, we start with the initial product state $|\Psi(t=0)\rangle=|0\rangle^{\otimes N}$ and compute its GGM ($\mathcal{G}$) at each  iteration step of the circuit.  The number of random realizations of the circuit considered here is $2\times 10^2$.  The behavior of the multipartite entanglement averaged over all such random realizations of the circuit with the iteration number $t$  is depicted in  Fig.~\ref{GGM_plot}.  From the figure, we note that $\mathcal{G}$ grows very fast and saturates eventually to a high value for large iteration times. We denote the saturation value (up to the  third decimal place of $\mathcal{G}$)  of multipartite entanglement   by $\mathcal{G}_{sat}$ and the iteration step require to reach the saturation by $t_{sat}^{\mathcal{G}}$.  For $N=12$, $\mathcal{G}_{sat}=0.478$ and $t_{sat}^{\mathcal{G}}=20$. Subsequently, in order to find an  approximate analytical form of $\mathcal{G}(t)$, we fit $\mathcal{G}/\mathcal{G}_{sat}$ (see the inset (a) of Fig.~\ref{GGM_plot}) and find that ${\mathcal{G}}$ grows approximately as
\begin{eqnarray}
{\mathcal{G}(t)} =\mathcal{G}_{sat}\tanh\left(\frac{t}{t_0}\right),
\label{GGM_growth}
\end{eqnarray}
with $t_0 \approx6$ for $N=12$.  The fitting is not exact and only provides an approximate  functional form  of $\mathcal{G}(t)$. We provide an estimate of the error involved in the fitting  and  the dependence of the constant $t_0$ on the system size $N$ in the Appendix \ref{appA}. A scaling analysis of $\mathcal{G}_{sat}$ with $N$ is also presented in the inset (b) which indicates that even for moderate   system size, $N=12$, the multipartite entanglement content of the random  state at moderately high iteration number eventually becomes very close to the maximum value of $\mathcal{G}$ in qubit systems.

\begin{figure}[t] 
\includegraphics[width=6cm]{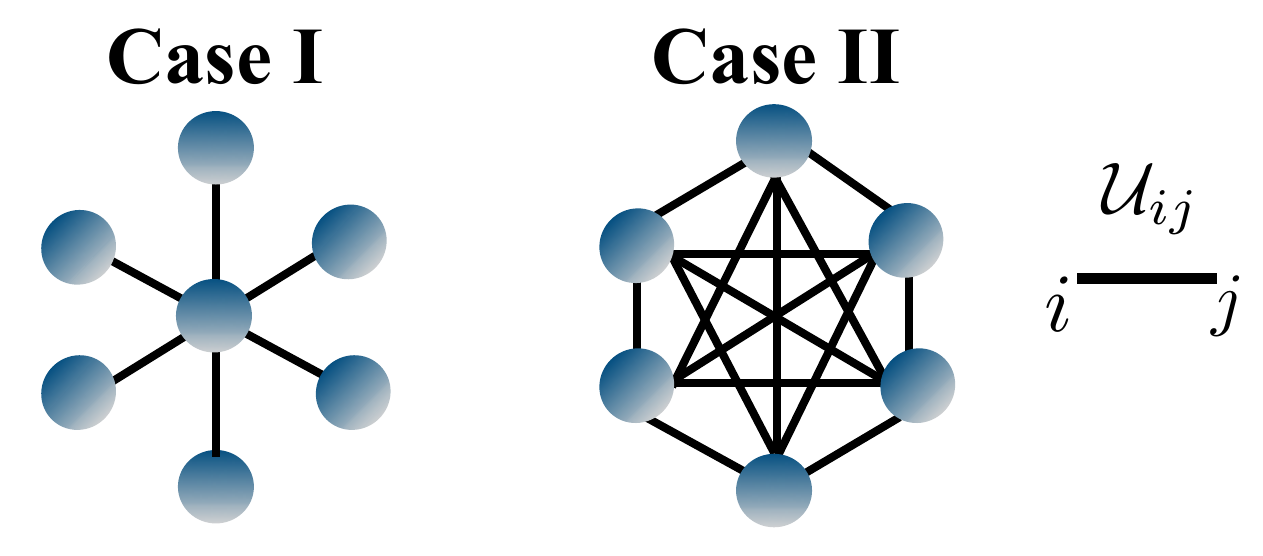} 
\caption{Schematic representation of different configurations of random unitary circuits. Case I illustrates the quasi long-range circuit where one qubit is connected with the rest of the  qubits. Case II depicts a complete long-range circuit where each  qubit is connected with all the other qubits. The solid black line represents the unitary $\mathcal{U}_{ij}$ acting on the qubits $i$ and $j$ at its two edges.} 
\label{GGM_star} 
\end{figure}

In addition to this, we mention here that the circuit configuration in Fig.~\ref{schematic} is the optimal one, in the sense that if we consider other variants of the circuit, where instead of short-range unitaries, the circuit comprised of unitaries acting on non-nearest neighbor qubits, 
there is no advantage of the multipartite entanglement generated at high iteration, as in that limit, all of them become ergodic and converge to the same Haar measure. However,  we observe that the rate of growth of genuine multiparty entanglement depends on the range and number of random unitaries considered. We elaborate this by considering two cases as follows (see Fig. \ref{GGM_star} for a schematic illustration of the cases).
In Case I, we consider a quasi long-range unitary circuit, such that two-qubit unitaries are now acting on the first qubit and rest of the qubits, i.e., $\Pi_{r=2}^{N}\mathcal{U}_{1r}$. Here, the number of unitaries acting at each iteration step remains the same as the previous case, which is $N-1$. We observe that in this case, for $N=12$,  to reach the saturation value  of multipartite entanglement the number of circuit iteration required is $t_{sat}^{\mathcal{G}}=11$. This implies,  by increasing the range of interaction the growth rate of multipartite entanglement can be increased even keeping the same number of unitaries as before.  Case II demonstrates a  proper long-range scenario in the sense that all the $N$ qubits are now connected through the two-qubit unitaries, $\Pi_{i<j, i, j=1}^{N}\mathcal{U}_{ij}$. The number of unitaries acting in the circuit at each iteration step is given by $\frac{N}{2}(N-1)$. We observe in this case, the growth rate is maximum and for $N=12$, the saturation occurs at  $t^{\mathcal{G}}_{sat}=3$. \\
  
\begin{figure}[h] 
\includegraphics[width=8.5cm]{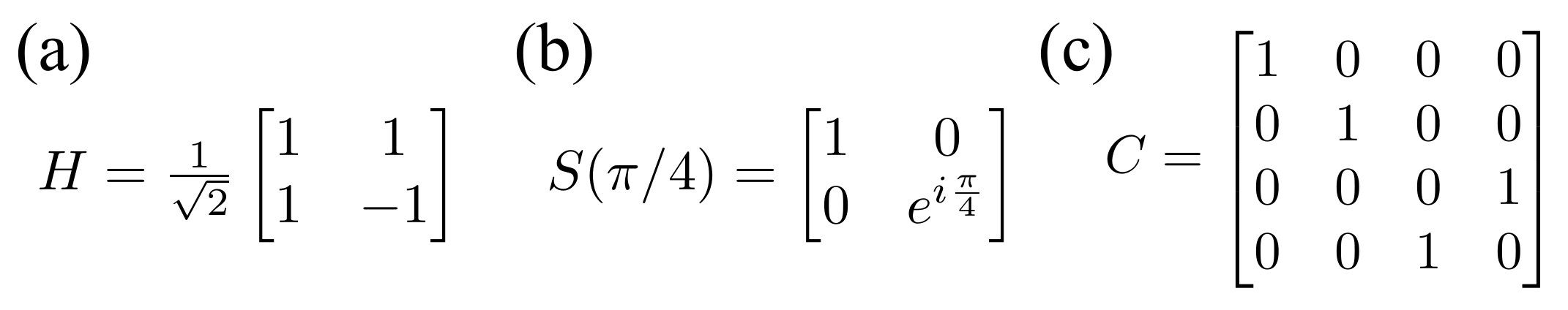} 
\caption{Elementary quantum gates of the Clifford group,  namely,  (a) the Hadamard gate, (b) the $\pi/4$ gate S, and (c) the controlled-NOT gate $C$, expressed in the computational basis.}
\label{clifford_gate_matrix} 
\end{figure}

\begin{figure}[h]
\includegraphics[scale=.32]{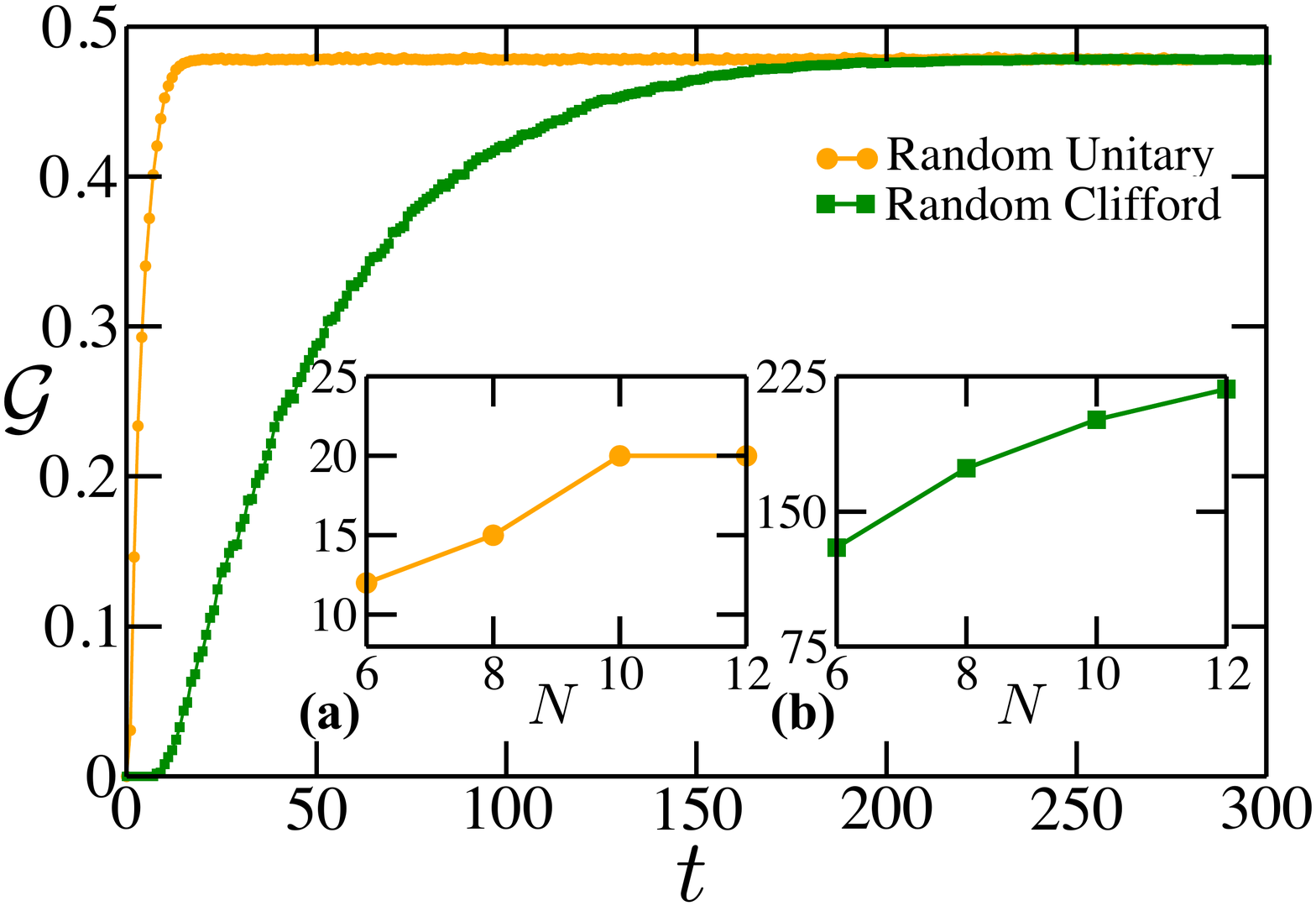}
\caption{Growth of genuine multiparty entanglement obtained for the random Clifford circuit (green squares). For a better comparison, we again plot the behavior of $\mathcal{G}$ obtained for the random unitary circuit (orange circles).  Here $N=12$ and in both the cases  averaging of $\mathcal{G}$ is performed over $2\times 10^2$ number of random realizations of  the circuits. In the insets, we compare the scaling of the iteration number ($t^{\mathcal{G}}_{sat}$)  required for the saturation of $\mathcal{G}$ in both (a)  random unitary circuits  and (b) random Clifford circuit with  system size $N$.} 
\label{Fig_clifford}
\end{figure}

We next consider another  example of a random quantum circuit, namely, the random Clifford circuit,  constructed by picking randomly any of the following gates with equal probability at each iteration of the circuit, a) Hadamard gate,  b) $S(\frac{\pi}{4})$,  and c) controlled-NOT \cite{Clifford1,Clifford2}. See Fig. \ref{clifford_gate_matrix}  for the matrix form of these gates.  In case of bipartite entanglement, it is known that though the  Clifford circuit can generate states with the same maximal entanglement entropy as Haar random states, the entanglement spectrum of such states is either flat or Poisson distributed which is different from the Wigner-Dyson distribution for the  Haar random states  \cite{Clifford3,Clifford4}. 
  Here we wish to see how distinct the behavior of multipartite entanglement remains for the random Clifford case than the random unitary scenario.  Fig. \ref{Fig_clifford} depicts the growth of genuine multipartite entanglement of the same initial product state $|\Psi\rangle=|0\rangle^{\otimes^N}$ that we have considered earlier, with each iteration step of the random Clifford circuit  (green squares). From the figure,  we can observe that in this case, multipartite entanglement grows with a relatively slower rate in comparison to the random unitary circuit case. The  reason for such behavior is that among the considered quantum gates that belong to the Clifford group, only controlled-NOT is capable of producing entanglement. On the other hand, for the random unitary circuits, the two-qubit random gates are more capable of generating entanglement between the sites. 
 The behavior is consistent with that obtained in earlier work \cite{Clifford5}, where it is reported that although the universal set of gates in random Clifford circuit  is capable of scrambling initial product states, the degree of randomness at intermediate steps of such states remain lower than those generated by two-qubit Haar random unitary circuits. However, for large time, the saturated value of multipartite entanglement ($\mathcal{G}_{sat}$) becomes almost the same as that obtained for the random unitary scenario. A comparison of the scaling  of the iteration step ($t^{\mathcal{G}}_{sat}$) required for the saturation of  multipartite entanglement  in both (a) random unitary circuit and (b) random Clifford circuit    with the size of the system ($N$) is also presented in the insets of the Fig. \ref{Fig_clifford}.

\section{Comparison  to random Matrix Product States}
\label{GGM_MPS}

We now compare the growth of multipartite entanglement obtained for random unitary circuit models as discussed in the previous section,   to that obtained for  random matrix product states. Matrix product states (MPS) with fixed bond dimensions lie in a tiny corner of the total Hilbert space and  are often found to be an approximate ground state of local Hamiltonians \cite{MPS_bunch1,MPS_bunch2,MPS_bunch3}. As stated earlier,  the random unitary circuit models represent a less-structured model than the Hamiltonian systems, and the presence of randomness eventually pushes the quantum state to occupy a wider region within the Hilbert space.   Therefore, the behavior of multipartite entanglement in random MPS and its comparison with the random quantum state generated through the random unitary circuit model is an interesting case to explore.  In Ref.~\cite{Random_MPS}, it has been reported that a set of non-homogeneous random MPS and the set of uniformly distributed general random states yield the same average states.
Here, our aim is to explore whether any such similarity exists between these two differently constructed random states when the multipartite entanglement properties are considered.

 \begin{figure}[h] 
\includegraphics[width=8.8cm]{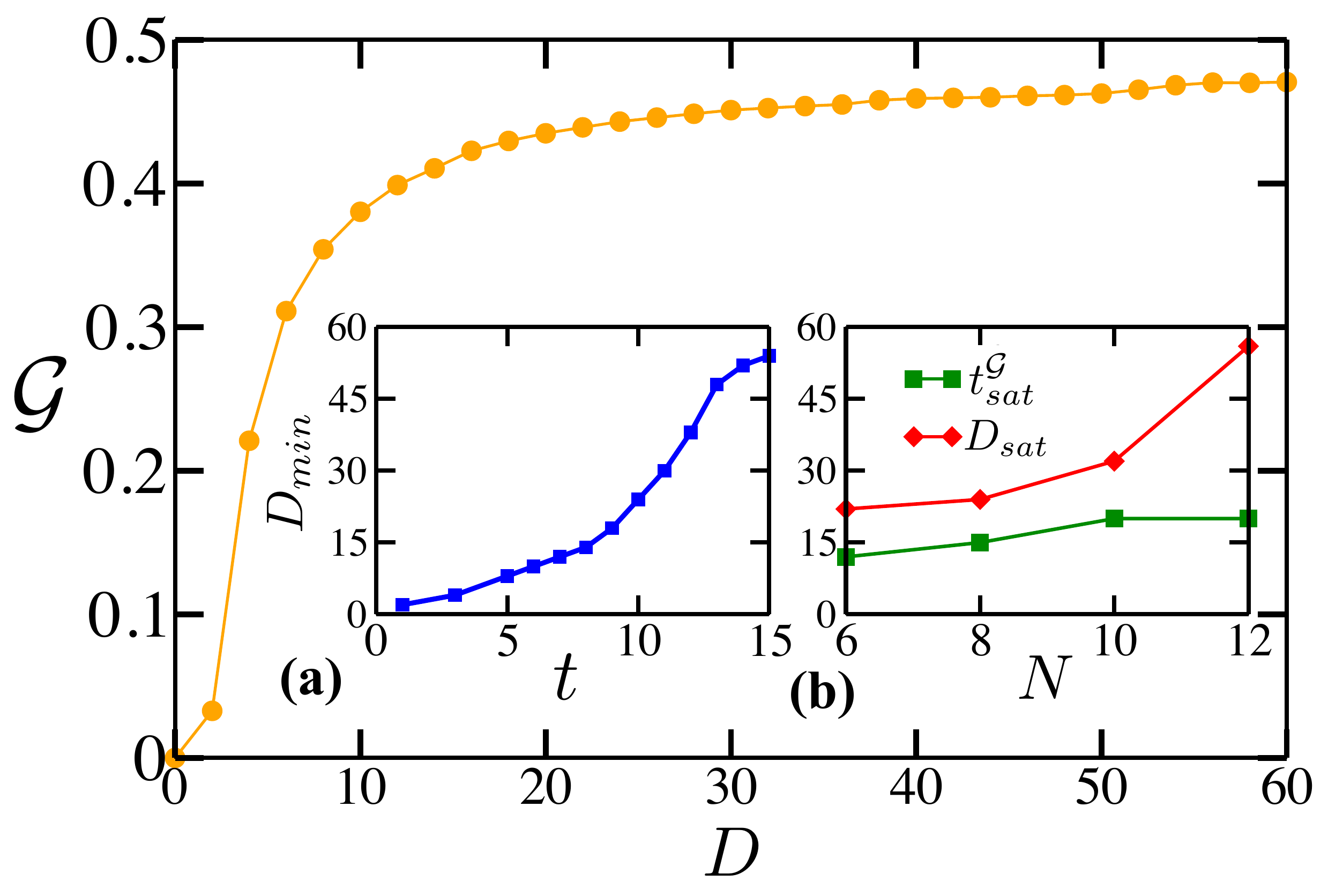} 
\caption{Growth of $\mathcal{G}$ in RMPS  with the bond dimension $D$.  Here, $N=12$ and the averaging of $\mathcal{G}$ is performed over $2\times 10^2$ random realizations. In the inset (a), we plot the minimum bond dimension $D_{min}$ required to reach the multipartite entanglement in a random MPS close to the value obtained after each iteration of the random unitary circuit.  The inset (b) depicts  the scaling $t^{\mathcal{G}}_{sat}$ and $D_{sat}$ for random unitary circuit and  random MPS respectively,  with the system size $N$.}
\label{fig:GGM_MPS}
 \end{figure}

We start with the matrix product states representation of any  pure quantum state $|\Psi\rangle$, which is given by 
\begin{eqnarray}
|\Psi\rangle=\sum_{i_1 i_2 \dots i_N} \text{Tr}(A_1^{i_1}A_2^{i_2}\dots A_N^{i_N}) |i_1 i_2 \dots i_N \rangle,
\label{Eqn:MPS}
\end{eqnarray}
where $A_k^{i_k}$ are  $D\times D$ complex matrices, with $D$ being the bond dimension. In general, in order to represent any quantum state in MPS form, one requires  $ND^2d$ number of parameters, where $d$ is the local Hilbert space dimension of the system (for qubits $d=2$). For small $D$, this number turns out to be much smaller than the dimension of the actual Hilbert space, $d^N$. 

 {
In this work, we consider a matrix product state, where the $A_k^{i_k}$ matrices are random unitaries $\mathcal{U}$ of dimension $D\times D$ 
and  aim to find the growth of the genuine multipartite entanglement with its bond dimension $D$.
Fig.~\ref{fig:GGM_MPS} shows  the growth of $\mathcal{G}$ of a random MPS with $D$.  For each $D$, averaging of $\mathcal{G}$ is performed over   $2\times 10^2$ random realizations of the matrix product states. We find that with increasing bond dimension, multipartite entanglement in RMPS increases and finally  saturates at a value $\mathcal{G}=0.470$, which is very close to that obtained for the random circuit. We denote  the bond dimension require to reach the saturation value by $D_{sat}$. For $N=12$, $D_{sat}=56$. Hence, we can argue that in terms of global entanglement content, the random quantum state generated at high iteration time of the circuit  is comparable to a random MPS with a moderately large bond dimension.

 A similar comparison for the points away from saturation is shown in the inset (a) of Fig. \ref{fig:GGM_MPS}, where we compute the minimum bond dimension ($D_{min}$) required for a random MPS to achieve multipartite entanglement close to that obtained after each iteration $t$   of the random unitary circuit. The closeness of the values of multipartite entanglement obtained in each of the cases has been quantified by the following factor $|\mathcal{G}^{RU}(t)-\mathcal{G}^{RMPS}(D_{min})|\leq 10^{-2}$. For example, the amount of multipartite entanglement obtained after $t=10$ iteration of the random unitary circuit is given by  $\mathcal{G}^{RU}=0.452$. Now for the random MPS, a value close to that, $\mathcal{G}^{RMPS}=0.443$ is obtained for $D=24$. Hence, we can argue that from the perspective of multipartite entanglement, the quantum state generated after $t=10$ circuit iterations is comparable to a random MPS with $D=24$.  At this stage, we would like to mention here that instead of choosing $A_i$ matrices as random unitaries $\mathcal{U}$, one can equally consider those as any general random matrices. However, we found that in that case, the amount of multipartite entanglement obtained for any bond dimension $D$ remains almost the same as that obtained for the unitary case (see Fig. \ref{MPS_non_unitary} in Appendix \ref{appB}). Therefore,  we can argue that choice of $A_i$ matrices as random unitaries turns out to be more beneficial as it involves relatively less number of random parameters than any general random matrix.
In addition to this, in the inset (b) of Fig.~\ref{fig:GGM_MPS}, we provide   a comparison of the scaling of  $D_{sat}$ obtained from random MPS and $t_{sat}^{\mathcal{G}}$  obtained from random unitary circuits, with the system size $N$.

\section{Delocalization of the wavefunction}
\label{GGM_IPR}
Along with the studies of the multipartite entanglement properties of the random quantum  circuits, we also study other global property of the models, namely,  the delocalization of an initial product state, when it evolves under the interactions of the random quantum gates. Importantly, this gives us an opportunity to compare the spread of wavefunctions in Hilbert space with the spread of entanglement in different bipartitions of a multipartite quantum state. In order to quantify the degree of delocalization of any quantum state in many-body Hilbert space, we consider a commonly used measure, known as the Inverse Participation Ratio (IPR) \cite{IPR_new,IPR_EE0}, for a given basis $\{|i\rangle\}_{i=1}^{2^N}$ which can be expressed as
\begin{eqnarray}\mathcal{I}=\frac{1}{\sum_{i=1}^{2^N} |\langle \Psi(t) |i\rangle|^4}.
\label{eq:IPR}
\end{eqnarray}
The measure IPR  is conventionally used to quantify the localization of any single-particle wavefunction in real space, which is attributed as the scenario of Anderson localization. However, the definition of localization is different in the context of many-body systems as in this case, the relevant space is the many-body Hilbert space rather than the real space of a single particle. To make a proper generalization of such concept to the many-body scenario, the basis states $\{|i\rangle\}_{i=1}^{2^N}$  considered here are such that the spatial regions of the lattice are not entangled with each other. In other words, each of such basis states is a fixed real-space configuration. An example of one such basis states, which we consider in this work  is the product basis of the local operator, e.g., $S_i^z$, $\{|0\rangle, |1\rangle\}^{\otimes N}$. However, choice of such a local basis is not unique and the value of $\mathcal{I}$ depends on the local basis considered, which can be further connected to the  quantum coherence of the state \cite{IPR_coherence1,IPR_coherence2}.\\

In general, for any many-body  quantum state $|\Psi\rangle_N$ completely delocalized in a given local basis $\{|i\rangle\}_{i=1}^{2^N}$}, one gets $\langle i|\Psi\rangle_N=\frac{1}{\sqrt{2^N}}, \forall i$, implying that $\mathcal{I}=2^N$. On the other hand, for a completely localized state, we have $\mathcal{I}=1$. In both  limits, multipartite entanglement becomes zero. However, there could be other states e.g., $|\psi\rangle_{\mbox{\small{GHZ}}}$,  for which  $\text{IPR}$ is very low but the multipartite entanglement content is maximum, i.e. $\mathcal{G}=0.5$. Therefore, for any generic quantum system, the exact relation between delocalization of the wavefunctions and its multipartite entanglement content is not  obvious.   In this regard,  there are works where attempts have been made to provide a relation between bipartite entanglement, quantified by measures such as averaged concurrence \cite{IPR_EE001}, purity \cite{IPR_EE001}, entanglement entropy \cite{IPR_EE1,IPR_EE2,IPR_EE3}, etc.,     and delocalization properties   quantified by Renyí entropy \cite{IPR_EE001}, participation ratio \cite{IPR_EE1,IPR_EE002,IPR_EE2,IPR_EE3},  for  certain quantum many-body wavefunctions.
   \begin{figure}[t]
  \includegraphics[width=8.8cm]{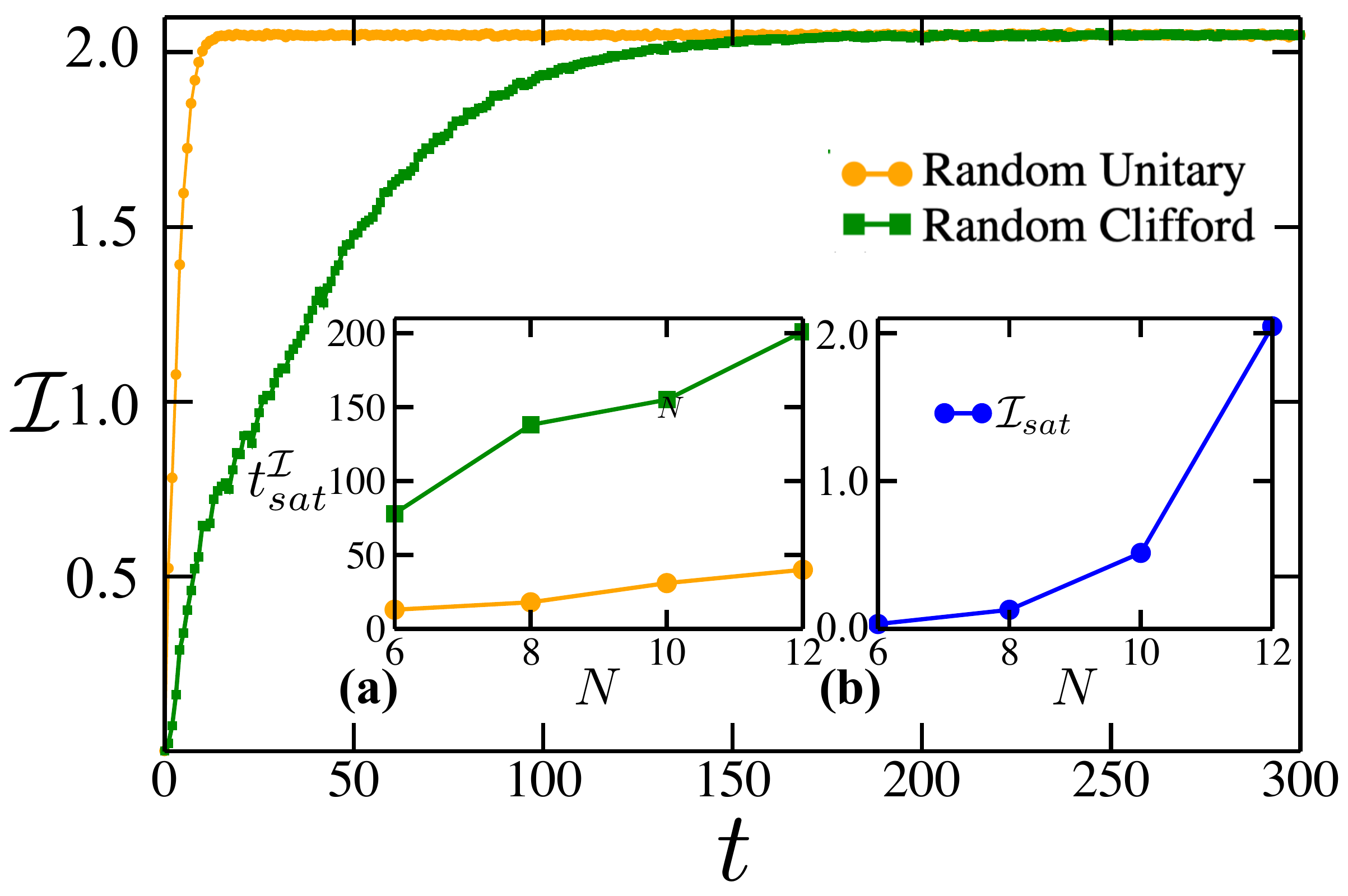}
  \caption{
  The growth of IPR ($\mathcal{I}$) with the iteration number $t$  for the random unitary circuit (orange circles) and random Clifford circuit (green squares). In both the cases, $2\times 10^2$ random realizations of the circuits have been considered. Here,  $N=12$.  In the inset (a), we compare the  scaling of iteration steps ($t_{sat}^{\mathcal{I}}$)  required to reach the saturation value of IPR in both random unitary and random Clifford circuits with system size ($N$).  Inset    (b) depicts the scaling of the saturation value of IPR ($\mathcal{I}_{sat}$)  with the system size $N$.  The $y$-axis of the main figure and  inset (b)  has been rescaled by $y=y\times 10^{-3}$.}
  \label{IPR_plot}
  \end{figure}
To investigate the relation between delocalization and growth of multipartite entanglement in both the random unitary   circuit and the random Clifford circuit, we again start with the  initial state $|\Psi(t=0)\rangle=|0\rangle^{\otimes N}$  and compute its IPR  at each iteration of the circuits. 
Fig.~\ref{IPR_plot} illustrates the growth of IPR ($\mathcal{I}$) with the iteration number $t$ for both random unitary circuit (orange circles) and random Clifford circuit (green squares). The averaging of IPR is performed  over $2\times 10^2$ random realizations of the circuits. We denote the saturation value (up to the second decimal place of $\mathcal{I} \times 10^{-3}$) of IPR    as $\mathcal{I}_{sat}$ and the number of circuit iteration require  to reach the saturation by $t^{\mathcal{I}}_{sat}$. 
Clearly, this behavior is qualitatively similar to the behavior obtained for multipartite entanglement in these circuits, as shown in Fig.~\ref{Fig_clifford}. Therefore, from the comparison, we argue that for the two different kinds of random circuits we consider in our work, the spreading of wavefunctions in Hilbert space and the growth of multipartite entanglement demonstrate a close correspondence. The inset (a) of Fig. \ref{IPR_plot} shows the comparison between the scaling of $t^\mathcal{I}_{sat}$ obtained for random unitary circuit and random Clifford circuit with the system size $N$. Along with this, inset (b) depicts the scaling of saturation value of IPR, $\mathcal{I}_{sat}$ with $N$.

We have now  characterized the properties of the random quantum state generated through the dynamics of the random unitary circuits. However, along with the generation of such highly entangled random quantum states,   to use it as a resource for quantum information processing and computational tasks, it is equally important to investigate its robustness properties. In the next section, we consider one such set-up and discuss its robustness properties in detail. 

\section{Robustness of multipartite entanglement}
\label{Weak_measurement}
Apart from its fundamental importance, multipartite entangled state has been found to be an useful resource in many quantum information and computation tasks \cite{multiparty_QIP1,multipartite_review,error_corr0,error_corr1}. 
In the case of measurement-based quantum information and computation schemes, starting from a highly entangled multipartite resource state, sequential measurements are applied to exploit the shared multipartite entanglement in accomplishing the desired tasks. 
In this way, the initial resource  is irreversibly degraded as the computation proceeds and reusability of the resource ceases. One way to minimize the effects of such quantum measurements is to apply the weak measurement schemes~\cite{weak_measure1,weak_measure2,weak_measure3,Anindita_saptarshi,Anindita_malda}. Though this, in turn, may affect the efficiency of the protocol, the reusability of the resource opens up. 
Another  example where one  is interested in doing certain  quantum computation tasks while keeping the multipartite entanglement is quantum error correction  \cite{error_corr0,error_corr1}. The protection of quantum information from the error introduced due to unavoidable interactions with the environment is the main aim of any quantum error-correcting codes.  In this regard, a key step is an efficient encoding process that maps the physical qubits into the encoded multipartite entangled logical state. This mapping process of the physical qubits to large Hilbert space essentially provides a scope to detect and even further correct errors of the physical qubit without destroying the logical state.  In general, the encoding process is comprised of the application of quantum gates on the physical qubit and the space of ancillary qubits. One such example is the encoding of an arbitrary qubit $|\psi\rangle=\alpha|0\rangle+\beta |1\rangle$ to the following multipartite entangled state by employing quantum circuit $U$ consisting of CNOT gates  $|\psi\rangle \xrightarrow[]{\text{U}} |\psi\rangle_{Encoded}=\alpha|0\rangle_L+\beta |1\rangle_L$, with $|0\rangle_L=|00\dots0\rangle$ and $|1\rangle_L=|111\dots 1\rangle$. 
In this respect, the usefulness of the random Clifford circuit as an efficient encoder for good quantum error-correcting codes has been studied in Ref.  \cite{error_corr2}. 
 Now  once the initial quantum bits are encoded,  the error is detected and often corrected by  performing  quantum measurements. One such example is the stabilizer measurement \cite{error_corr0,error_corr1}. However, schemes for quantum-error correction that employ feedback and weak measurement have also been proposed \cite{error_corr3,error_corr4}.  All these studies motivate us to examine the robustness of the quantum state generated at large iteration times of the circuit under the effect of weak quantum measurements, which essentially paves the path for the next step: designing an efficient error-correcting code using the multipartite entangled state generated through the random unitary circuit dynamics. However, such an investigation demands separate attention which is beyond the scope of the current work and we wish to explore that elaborately in our future work.

\begin{figure}[h]
\includegraphics[width=8.cm]{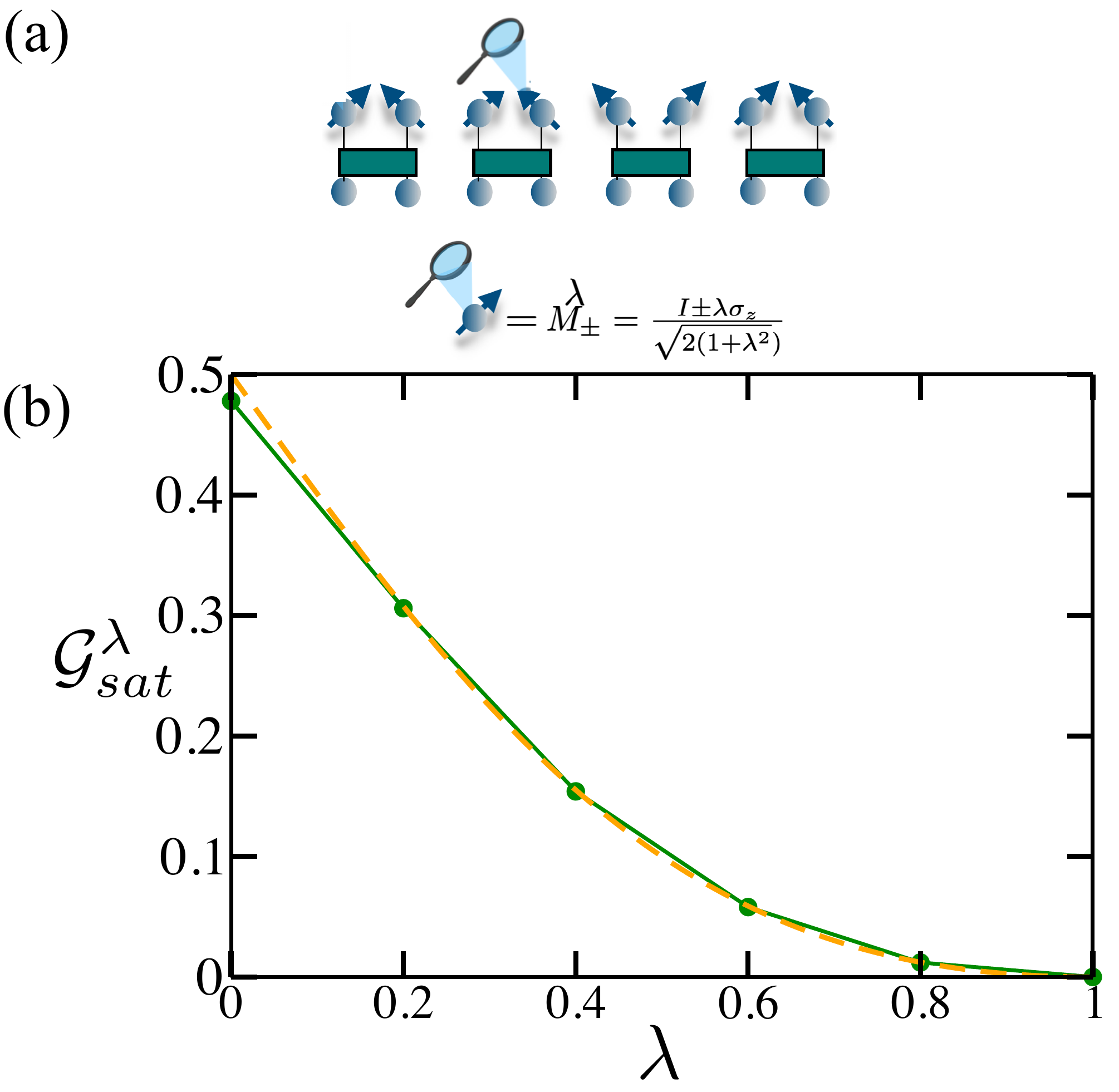}
\caption{Depiction of weak measurement settings. Panel (a) represents a schematic sketch of the weak measurements  $M_{\pm}^{\lambda}$ performed on any qubit randomly chosen from the $N$ sites. Panel (b) displays the decreasing behavior of saturation value of multipartite entanglement $\mathcal{G}_{sat}^{\lambda}$ with   measurement strength $\lambda$. The dashed orange line signifies the fitted  analytical expression given in Eq.~(\ref{natok}). Here, $N=12$ and the number of random realizations of the circuit  is $2\times 10^2$. 
}
\label{ggm_weak}
\end{figure}

We consider weak measurement operators,  $M^{\lambda}_{\pm}=\frac{I\pm\lambda\sigma_z}{\sqrt{2(1+\lambda^2)}}$, characterized by the parameter $0 < \lambda \leq 1$. These measurement operators satisfy  the completeness relation: $M_+^{\lambda} M_+^{\lambda \dagger}+M_-^{\lambda} M_-^{\lambda \dagger}=\mathbb{I}$. The parameter $\lambda$ represents the measurement strength. Indeed, for $\lambda=1$, $M_{\pm}^\lambda$ correspond to projective measurements. We perform $M_{\pm}^\lambda$ on the quantum state generated at large iteration  time ($t=50$)  of the random unitary circuit and the position of such measurements is completely random. For a schematic depiction, see  Fig.~\ref{ggm_weak}(a).  The  quantum state after performing the weak measurement is then reads as  $|\Psi(t)\rangle_{\pm}^{\lambda}=\frac{M_{\pm}^{\lambda}|\Psi(t)\rangle}{||M_{\pm}^{\lambda}|\Psi(t)\rangle||}$. The procedure is repeated for a large number of measurements ($10^2$) and the final value of the global entanglement is obtained by averaging over all such outcomes and  a large number of random realizations of the circuits ($2\times 10^2$).
The behavior of the saturated value of global entanglement, denoted by $\mathcal{G}_{sat}^\lambda$, with the measurement strength $\lambda$ is depicted in Fig.~\ref{ggm_weak}(b). From the figure, we find that the  multipartite entanglement of the random state obtained at high iteration time exhibits a polynomial decay with the strength of measurement, which can be analytically expressed as
\begin{equation}
\mathcal{G}_{Sat}^\lambda\approx\frac{(1-\lambda)^2}{2(1+\lambda^2)}.
\label{natok}
\end{equation}
Interestingly, the above analytical form coincides exactly with the decay profile of multipartite entanglement of an $N$-qubit pure GHZ state.

\section{Conclusion}
\label{conclusion}
In this work, we analyzed the global quantum properties of the random quantum circuits, which are generally considered as least-structured models for quantum dynamics. We considered two structurally different kinds of random quantum circuits, namely, random unitary circuits comprised of short-range, and long-range Haar uniformly generated unitaries and random Clifford circuits and studied the growth of genuine multipartite entanglement when an initial product state is iteratively subjected to those circuits. We observed that for random unitary circuits, the initial product state accumulates a high amount of multipartite entanglement even after a few iterations of the circuits. However, the growth rate is relatively slow in the case of a random Clifford circuit, and in this case, to reach the same saturation value, a large number of circuit iteration is required.   In recent times, there have been both theoretical proposals \cite{ref24,ref25} and experimental developments \cite{multiparty_ent2,multiparty_ent6,multiparty_ent7,multiparty_ent8} on controlled preparation of highly multipartite entangled states. In that respect, our results propose a scheme for generation of a highly multipartite entangled state from a relatively  simpler and less-structured model, which can be a promising scheme  for efficient  generation and control of multipartite entangled state.  We then compared the behavior of global entanglement obtained for random unitary  circuits to that of a random matrix product states. We report that the behavior of genuine multipartite entanglement is very similar in both cases. In particular, we observed that a random matrix product state with a  moderately high bond dimension attains the value of genuine multipartite entanglement close to that obtained for the random state generated for large iteration of the random unitary circuits.\

 In addition to this, we made a connection between the behavior of multipartite entanglement with the other global properties of the system, such as the delocalization of the initial wavefunctions in Hilbert space and observed a very close correspondence between these two global quantities.  In both the random quantum circuits, the qualitative behavior of delocalization measure remains very much akin to the multipartite entanglement obtained in those circuits.  Finally, we studied the robustness of the multipartite entanglement generated through such random unitary dynamics, under the effect of weak measurements performed on any qubit, randomly chosen from $N$ sites. We showed that the circuit sustains a non-zero amount of global entanglement even when the strength of the measurement is very high.  The analysis is  arguably a step towards   the design of an efficient scheme such as  quantum error-correcting code, measurement-based quantum computation, etc., using the multipartite entangled state generated through the random unitary circuit dynamics, which we aim to explore in our future works.

\section*{Acknowledgements}
The Authors thank H. S. Dhar, U. Sen,  J. Rodr{\'i}guez Laguna,  G. Sierra and A. Misra, for providing their fruitful suggestions. S.S.R. acknowledges Ministerio de Ciencia, Innovaci{\'o}n y Universidades (Grant No. PGC2018-095862-B-C21), Comunidad de Madrid (Grant No. QUITEMAD+ S2013/ICE-2801), Centro de Excelencia Severo Ochoa Programme (Grant No. SEV-2016-0597), and CSIC Research Platform on Quantum Technologies PTI-001.
\\
\\
\appendix

\section{Approximate fitting of $\mathcal{G}/\mathcal{G}_{sat}$ vs $t$ plot}
\label{appA}
\begin{figure}[h]
\includegraphics[scale=.35]{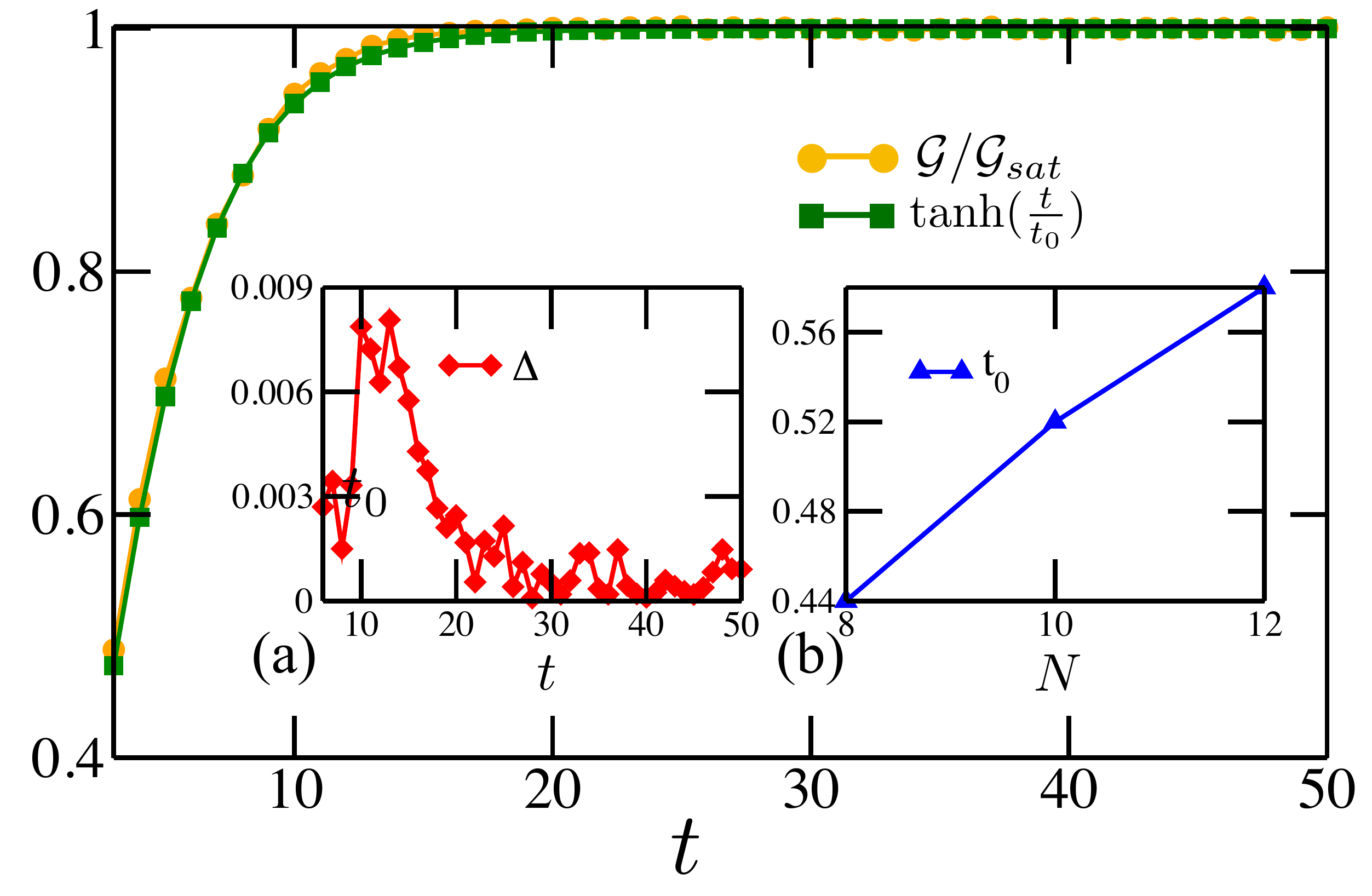}
\caption{Fitting of $\mathcal{G}/\mathcal{G}_{sat}$ in random unitary circuit by $\tanh(\frac{t}{t_0})$ with $N=12$. Inset (a) shows the variation of error $\Delta=|\mathcal{G}(t)/\mathcal{G}_{sat}-\tanh(\frac{t}{t_0})|$ for the whole region of $t$ we considered. Inset  (b) depicts the scaling of $t_0$ with N.}
\label{tanh_fit}
\end{figure}

In this section, we provide an estimate of the error introduced due to approximation of $\mathcal{G}(t)/\mathcal{G}_{sat}$ as a near functional form $\tanh(\frac{t}{t_0})$. We provide the fitting in  Fig. \ref{tanh_fit}. From the inset (a) of the figure, we can see that the error $\Delta=|\mathcal{G}(t)/\mathcal{G}_{sat}-\tanh(\frac{t}{t_0})|$  never exceeds the value $9\times 10^{-3}$ and for high $t$ ($t>18$), it  even becomes $\Delta< 3\times 10^{-3}$. The scaling of the constant $t_0$ with $N$ is depicted in the inset (b). 

\section{Random matrix product states with any general random $A_i$ matrices}
\label{appB}
\begin{figure}[h]
\includegraphics[scale=.36]{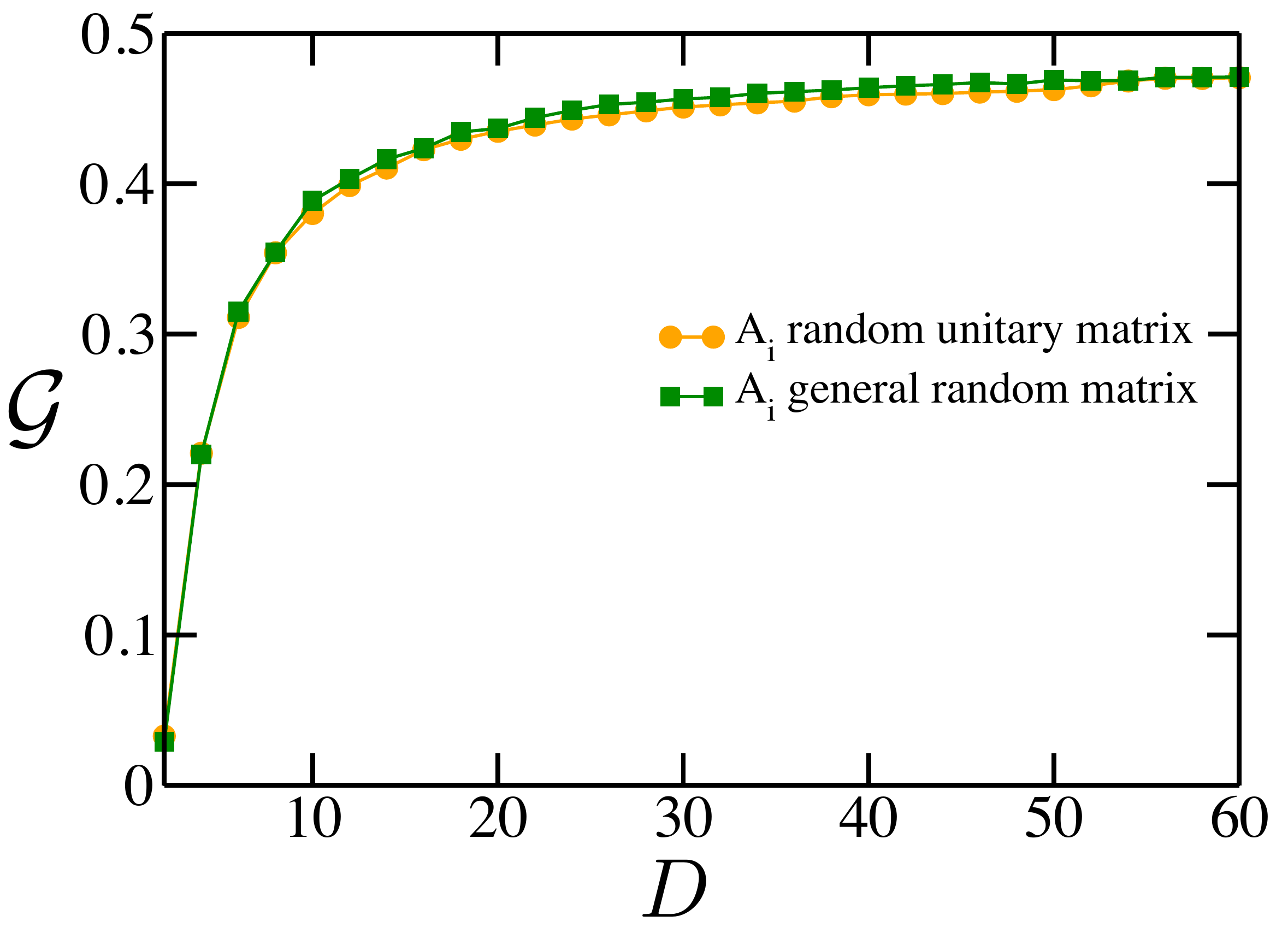}
\caption{Comparison of the growth of genuine multipartite entanglement as quantified by $\mathcal{G}$ with the bond dimension of the $A_i$ matrices $D$, when  $A_i$ are chosen as random unitaries   $\mathcal{U}$ (orange circles) and any general random matrix (green squares). Here, number
 of random realizations of the $A_i$ matrices have been considered is $2\times 10^2$ and  $N=12$.} 
\label{MPS_non_unitary}
\end{figure}

In this section, we provide a comparison of the multipartite entanglement obtained by considering the $A_i$ matrices in the random MPS as given in Eq. (\ref{Eqn:MPS}) of the main text, as $D\times D$ random unitaries and any general $D\times D$ random matrix without the unitary constrain. Fig. \ref{MPS_non_unitary} shows that in both cases, the value of multipartite entanglement for any $D$ remains almost the same. Hence, in terms of the number of random parameters, the choice of $A_i$ matrices as random unitaries will be beneficial as it consists relatively less number of random parameters. 

\end{document}